**Galileo ionosphere profile coincident with repeat plume detection location at Europa**

Melissa A. McGrath[1] and William B. Sparks[2]
[1]SETI Institute, Mountain View, CA mmcgrath@seti.org
[2]Space Telescope Science Institute, Baltimore, MD sparks@stsci.edu

Multiple plume detections at Europa have now been reported using the Hubble Space Telescope with independent techniques (Roth et al. 2014; Sparks et al. 2016). A repeat detection of a plume at the same location has also been reported (Sparks et al. 2017), and this location, near the well-known crater Pwyll, is coincident with a previously-reported thermal anomaly observed with the Galileo PPR in 1997 (Spencer et al. 1999). Figure 21a of Sparks et al. (2016) shows the locations of the four positive plume detections, which we designate as pR1, pR2, pS1, and pS2 in Figure 1(a), with pS1 being the repeat detection near Pwyll. Figure 2 of Sparks et al. (2017) shows the correspondence between the location of the repeat plume detection and the thermal anomaly.

We have discovered that the location of the repeat plume detection also nearly coincides with the Galileo radio occultation E6a entrance profile (Kliore et al. 1997) as illustrated in Figure 1(a), which shows the locations of all positive plume detections to date (open triangle symbols), as well as the locations of the 10 Galileo radio occultation ionosphere profiles (square symbols). The geometry of the two observations - the repeat plume detections and the E6a entrance occultation measurement - is virtually identical, with the off-limb enhancement observed at the dawn terminator, just as the region has moved from night to day. The compilation of all the Galileo radio occultation results was discussed by Kliore et al. (2006) and published in McGrath et al. (2009) Figure 7. It is included here as Figure 1(b), and shows that the E6a entrance profile (E6an) was the strongest of the 6 positive ionosphere detections made by Galileo. The peak electron density of the E6an detection was $\sim 1.3 \times 10^4 cm^{-3}$ at an altitude of ~50km, while Sparks et al. (2017) estimated the height for this plume to be ~50km. [By contrast, plumes pR1, pR2 and pS2 were roughly 200km in height.] The neutral gas column density corresponding to the E6an electron density was estimated by Kliore et al. (1997) to be $\sim 3 \times 10^{16} cm^{-2}$, while the plume gas column density corresponding to the repeat plume detection by Sparks et al. (2017) was estimated to be $\sim 1.8 \times 10^{17} cm^{-2}$, which is in reasonably good agreement given the qualitative nature of both estimates.

Until the detection of active plumes, the primary source for Europa's atmosphere was thought to be sputtering of surface water ice (e.g., Johnson et al. 1982). It has been challenging with such a source to explain the highly non-uniform ionosphere profiles and auroral emissions observed at Europa (e.g., Cassidy et al. 2007). Plumes provide a much more natural explanation for pronounced spatial inhomogeneities in both the atmosphere and ionosphere. This new result also illustrates that ionosphere measurements may provide another important means to detect active plumes.

The coincidence of four independent positive detections (pS1 observed twice, a thermal anomaly, and the strongest ionosphere profile) at the same location lends credence to the idea that activity at the near-Pwyll location may be long-lived, given the 17-19 year separation between the Galileo thermal and ionosphere measurements (both made in 1997), and the repeat plume detections in 2014 and 2016.

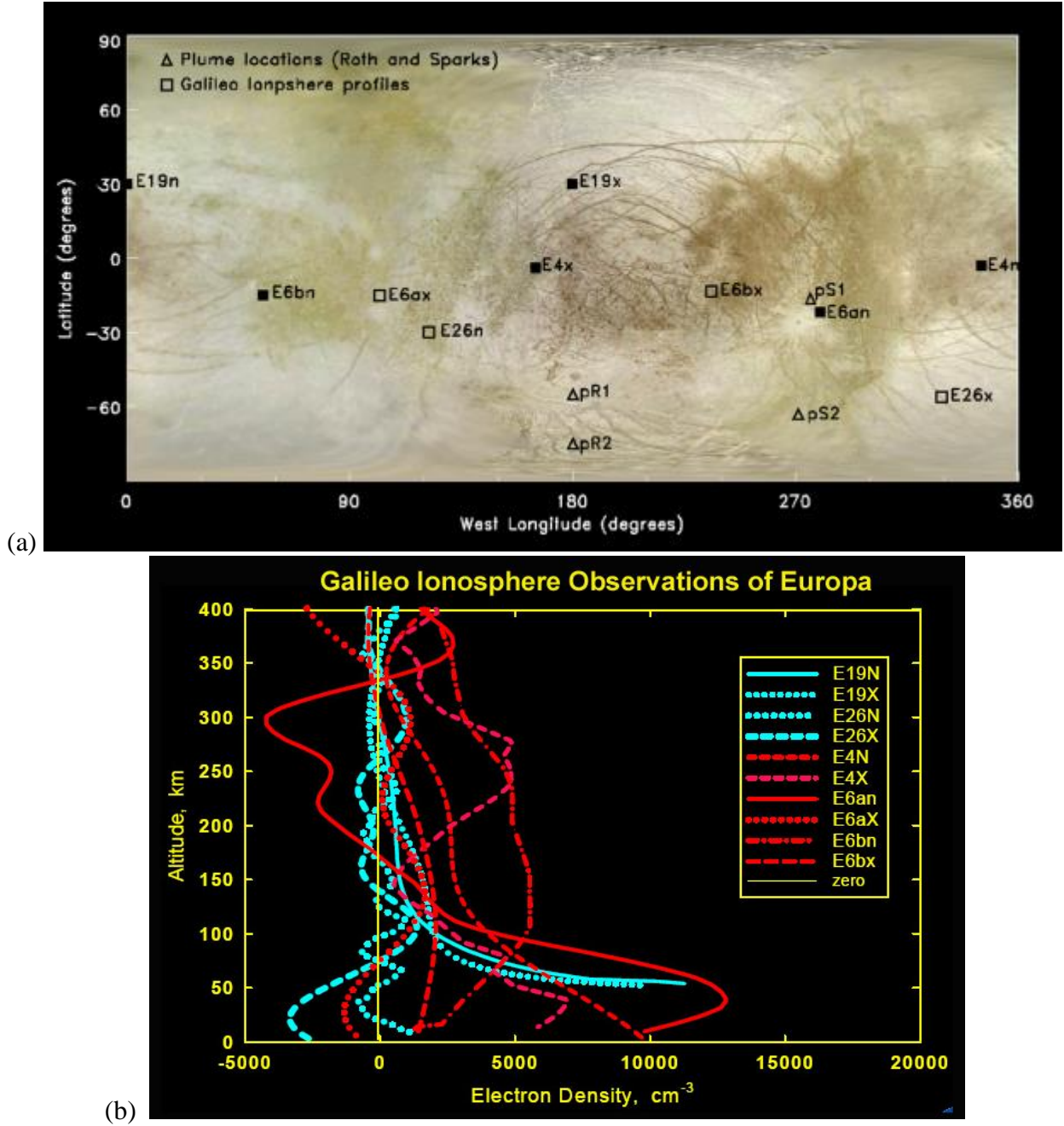

Figure 1. (a) The location on the surface of Europa corresponding to the Galileo radio occultation ionosphere profile measurements (square symbols) and the locations of plumes detected to date by Roth et al. (2014) and Sparks et al. (2016) (triangle symbols) using the Hubble Space Telescope. (b) The compendium of measured ionosphere profiles by Galileo radio occultation (Kliore et al. 1997; Kliore et al. 2006 - shown here is Figure 7 from McGrath et al. (2009)). Occultation entrance ionosphere profiles are indicated by an 'n', exit profiles by an 'x'. Positive detections are shown with filled square symbols; non-detections are shown with open square

symbols. The four plume detections reported by Roth et al. (2014) and Sparks et al. (2016) are located at ~180 W long, -55 and -75 S latitude (Roth plumes pR1 and pR2 in the figure) and ~275.7 W longitude, -16.4 S latitude (Sparks plume pS1 in the figure); ~271 W longitude, -63 S latitude (Sparks plume pS2 in the figure) - see Figure 21a of Sparks et al. 2016 for a polar projection map showing the location of all four plumes including uncertainty estimates. pS1 has been detected twice, once in March 2014, and once in February 2016. The only close correspondence between Galileo ionosphere measurements and plume detections is for profile E6an, which is located close to the repeat plume detection near Pwyll crater. E6an was the strongest of the 6 positive ionosphere detections, as shown in (b), with a peak electron density of ~$1.3 \times 10^4$ cm$^{-3}$ at an altitude of ~50km.

**References**

Cassidy, T. A. et al. 2007, Icarus, 191, 755
Kliore, A. et al. 1997, Science, 277, 355
Kliore, A. et al. 2006, presentation at 36th COSPAR Scientific Assembly, Beijing, China, Abstract #2599
Johnson, R. E. et al. 1982, Nucl. Instr. Meth., 198, 147
McGrath, M. A. et al. 2009, Chapter 21 in *Europa*, Eds. R. Pappalardo, W. McKinnon, and K. Khurana, University of Arizona Press Space Science Series, p. 485
Roth, L. et al. 2014, Science, 343, 171, doi:10.1126/science.1247051
Sparks, W. B. et al. 2016, ApJ, 829, 121, doi:10.3847/0004-637X/829/2/121
Sparks, W. B. et al. 2017, ApJL, 839, L18, doi:10.3847/2041-8213/aa67f8
Spencer, J. R. et al. 1999, Science, 284, 1514